\documentclass[twocolumn,nofootinbib,showpacs,showkeys,prd,superscriptaddress]{revtex4}
\usepackage{graphicx}
\usepackage{amsmath}
\usepackage{bm}

\begin{document}

\title{Time-Domain Measurement of Broadband Coherent Cherenkov Radiation}
\date{\today}

\author{P.~Mio\v{c}inovi\'c}
\email{predrag@phys.hawaii.edu}
\affiliation{Department of Physics and Astronomy, University of Hawaii at Manoa, Honolulu, Hawaii}
\author{R.~C.~Field}
\affiliation{Stanford Linear Accelerator Center, Stanford University, Menlo Park, California}
\author{P.~W.~Gorham}
\affiliation{Department of Physics and Astronomy, University of Hawaii at Manoa, Honolulu, Hawaii}
\author{E.~Guillian}
\affiliation{Department of Physics and Astronomy, University of Hawaii at Manoa, Honolulu, Hawaii}
\author{R.~Milin\v{c}i\'c}
\affiliation{Department of Physics and Astronomy, University of Hawaii at Manoa, Honolulu, Hawaii}
\author{D.~Saltzberg}
\affiliation{Department of Physics and Astronomy, University of California at Los Angeles, Los Angeles, California}
\author{D.~Walz}
\affiliation{Stanford Linear Accelerator Center, Stanford University, Menlo Park, California}
\author{D.~Williams}
\altaffiliation[Present address: ]{Pennsylvania State University, State College, Pennsylvania}
\affiliation{Department of Physics and Astronomy, University of California at Los Angeles, Los Angeles, California}

\begin{abstract}
We report on further analysis of coherent microwave Cherenkov impulses emitted via the Askaryan
mechanism from high-energy electromagnetic showers produced at the Stanford Linear Accelerator 
Center (SLAC). In this report, the time-domain based analysis of the measurements made with 
a broadband (nominally 1-18 GHz) log periodic dipole array antenna is described. The
theory of a transmit-receive antenna system based on {\it time-dependent 
effective height} operator is summarized and applied to fully 
characterize the measurement antenna system and to reconstruct the electric field 
induced via the Askaryan process. The observed radiation intensity and phase as functions of frequency 
were found to agree with expectations from 0.75--11.5~GHz within experimental errors on the normalized
electric field magnitude and the relative phase; $\sigma_{R|E|}= 0.039\;\mu\text{V/MHz/TeV}$ and 
$\sigma_\phi=17^\circ$. This is the first time this agreement has been observed over such a broad bandwidth,
and the first measurement of the relative phase variation of an Askaryan pulse. The importance of 
validation of the Askaryan mechanism is significant since it is viewed as the most promising way to 
detect cosmogenic neutrino fluxes at $E_\nu\agt10^{15}$~eV.
\end{abstract}


\maketitle

\section{Introduction}

G.~Askaryan proposed in 1962 that the charge excess in a compact particle shower in a dielectric 
medium will produce a coherent radio Cherenkov emission~\cite{askaryan}. 
Subsequent theoretical work supported this prediction~\cite{zhs,am,razz}. The 
experimental verification came in 2001~\cite{slac01}, with follow up 
measurements confirming the frequency and the polarization properties of the 
emitted radiation~\cite{slac05}. 

The interest in the characterization of the Askaryan process comes from the idea that it can be used to detect 
ultra-high-energy astrophysical neutrinos ($E_\nu\agt10^{15}$~eV) interacting in radio
transparent dielectric media, {\it e.g.}, ice~\cite{rice,forte,anita-lite,anita}, salt~\cite{slac05}, or Moon's 
regolith~\cite{dag,glue}. Considering the increasing experimental effort to observe such neutrinos, it is of 
importance to understand the properties of coherent Cherenkov radiation by verification of 
theoretical models. Furthermore, it is of practical value to provide the
time-domain characterization of coherent Cherenkov radiation since the experimental triggering and, to 
large extent, the data analysis will be based on time-domain properties of the signal. 

The emission of a coherent radio signal comes from the charge asymmetry in particle shower development.
The asymmetry is due to combined effects of positron annihilation and Compton scattering on 
atomic electrons. There is $\sim$20\% excess of electrons over positrons in a particle shower, 
which moves as a compact bunch, a few cm wide and $\sim$1 cm thick, at a velocity above the speed of 
light in the medium. The frequency dependence of Cherenkov radiation emitted is $dP\propto\nu d\nu$. 
For radiation with wavelength $\lambda\gg l$, where $l$ is the length scale of the particle bunch, the 
radiated electric field will add coherently and thus be proportional to the shower energy. 
A radio signal emitted by a particle shower in a dielectric material such as ice or salt is coherent up to 
a few GHz, is linearly polarized, and lasts less than a nanosecond. A shower with 
energy of $10^{19}$ eV interacting in the ice will produce a radio pulse with peak strength of 
$\sim1$~mV/m/MHz at the distance of 1~km. 

This report describes the analysis of measurements of of Askaryan pulses recorded with a log periodic dipole
array (LPDA) antenna in an experiment (SLAC T460) performed at 
the Final Focus Test Beam (FFTB) facility at SLAC in June 2002~\cite{slac05}. The data taking is 
described first, followed by the discussion of a time-domain based description of the antenna system response to 
transient radiation, and the description of system calibration. In the last two sections, the
reconstruction of the electric field induced by the particle shower is presented and compared to 
the theoretical work. 

\section{Measurements}

\begin{figure*}[t]
\includegraphics[scale=0.5]{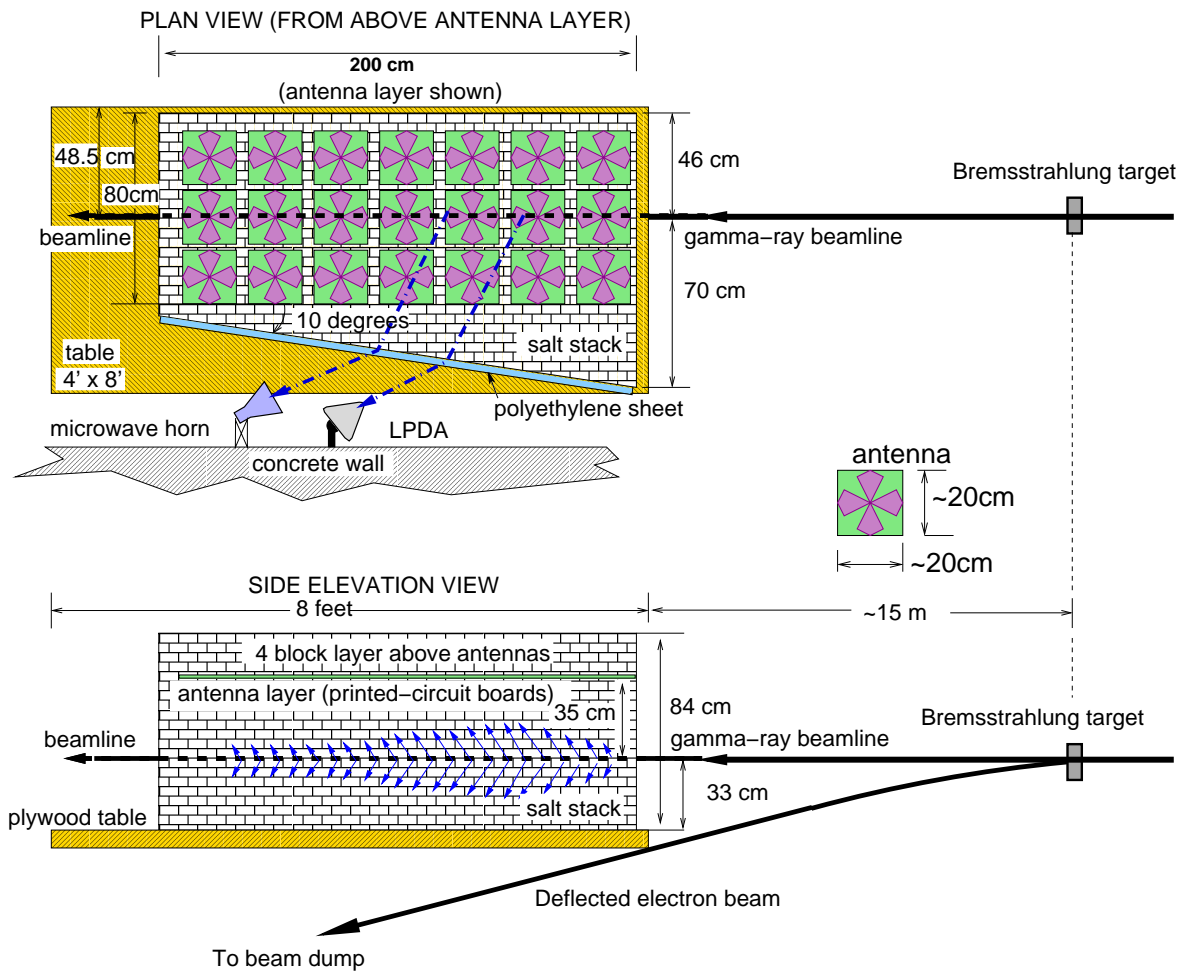}
\caption{Geometry of salt-block target and receiving antennas in SLAC T460 experiment, from Ref.~\cite{slac05}.
\label{fig:setup}}
\end{figure*}

The experimental setup is illustrated in Fig.~\ref{fig:setup}, which shows top and side views of the salt-block 
target, receiving antenna locations, and the incoming beam.
The particle showers in the salt target were induced by bremsstrahlung gamma-ray photon bunches emanating from an 
aluminum radiator plate placed in the path of the electron beam. The electron beam itself was deflected to pass below 
the salt target. The electrons in the beam were accelerated to $\sim$28.5~GeV and the energy of secondary bremsstrahlung 
photons followed $E^{-1}$ distribution up to $\sim$28.5~GeV. The total energy of photons in each bunch varied with the
electron beam intensity and the thickness of the radiator plate.

The salt target was constructed from salt bricks, with a total mass of about 4 metric tons. The detailed description of 
the target construction can be found in Gorham {\it et al.}~\cite{slac05}.
The radio pulses arising from photon induced particle showers were collected by bowtie antenna array embedded in the salt 
and by a C/X-band horn antenna and an LPDA antenna outside the salt. The bowtie antennas were located in a plane parallel 
and 
about 35~cm above the shower axis. The horn and LPDA antennas were placed in the plane of the shower, with LPDA pointing 
approximately at the expected shower maximum, as indicated in the Fig.~\ref{fig:setup}. The side of salt target facing the
external antennas was angled at $10^\circ$ in order to allow transmission of Cherenkov radiation which would otherwise
have been totally internally reflected. The antennas were read out using digital 
oscilloscopes, and to improve signal-to-noise ratio (SNR), the recorded waveforms were averaged over many pulses, 
using an ultra-stable microwave transition-radiation trigger from an upstream location.

The majority of data collected in the experiment was analyzed and published in Ref.~\cite{slac05}. In 
order to further test theoretical models of the coherent Cherenkov radiation,
this work will concentrate on the analysis of the data collected with the LPDA in two runs, 35 and 109, in which no 
microwave filters were placed between the LPDA and the oscilloscope,\footnote{Although, a 20~dB attenuator was used in run 
109 to restrict the peak voltage to the dynamic range of the oscilloscope.} providing the broadest bandwidth data. 
The total photon bunch energies in these runs were estimated to be 0.48 and 1.1~EeV, respectively, with an uncertainty on 
the order of 20\% due to the beam intensity and the electron bunch energy distribution fluctuations.

The LPDA is an antenna constructed by arranging multiple dipoles in a geometry that provides high 
signal gain over a large frequency bandwidth~\cite{kraus}. The LPDA antenna used was Electro-Metrics 
model EM-6952 with nominal bandwidth from 1--18~GHz. The antenna was connected by two pieces of 75--foot heliax 
cable, Andrew LDF4--50A, and by three pieces of 12--inch semi-rigid Haverhill cable, to a CSA8000 sampling oscilloscope 
with 20~GHz bandwidth and up to 1000~GSa/s sampling rate.

\begin{figure}[t]
\includegraphics[scale=0.45]{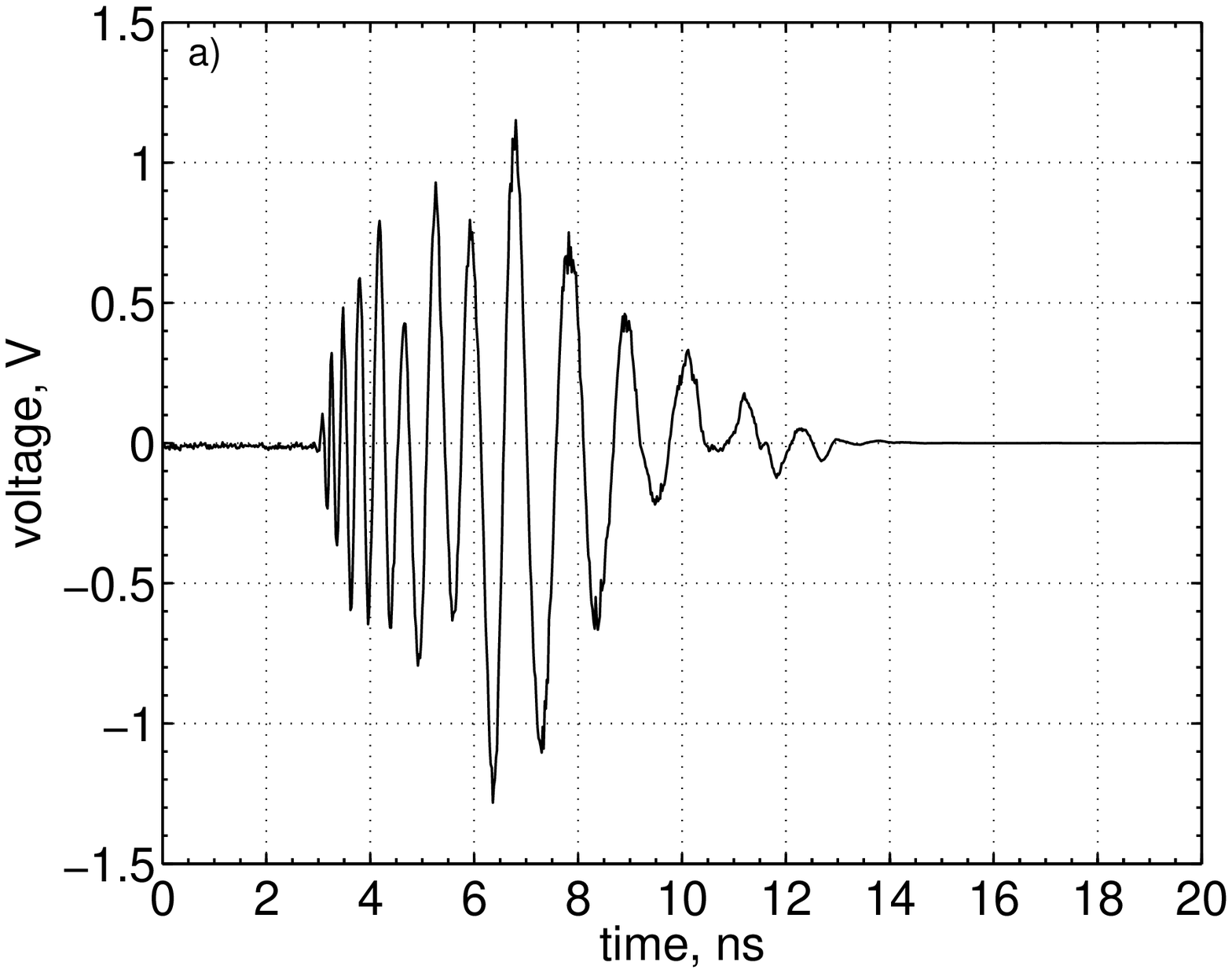}\\
\includegraphics[scale=0.45]{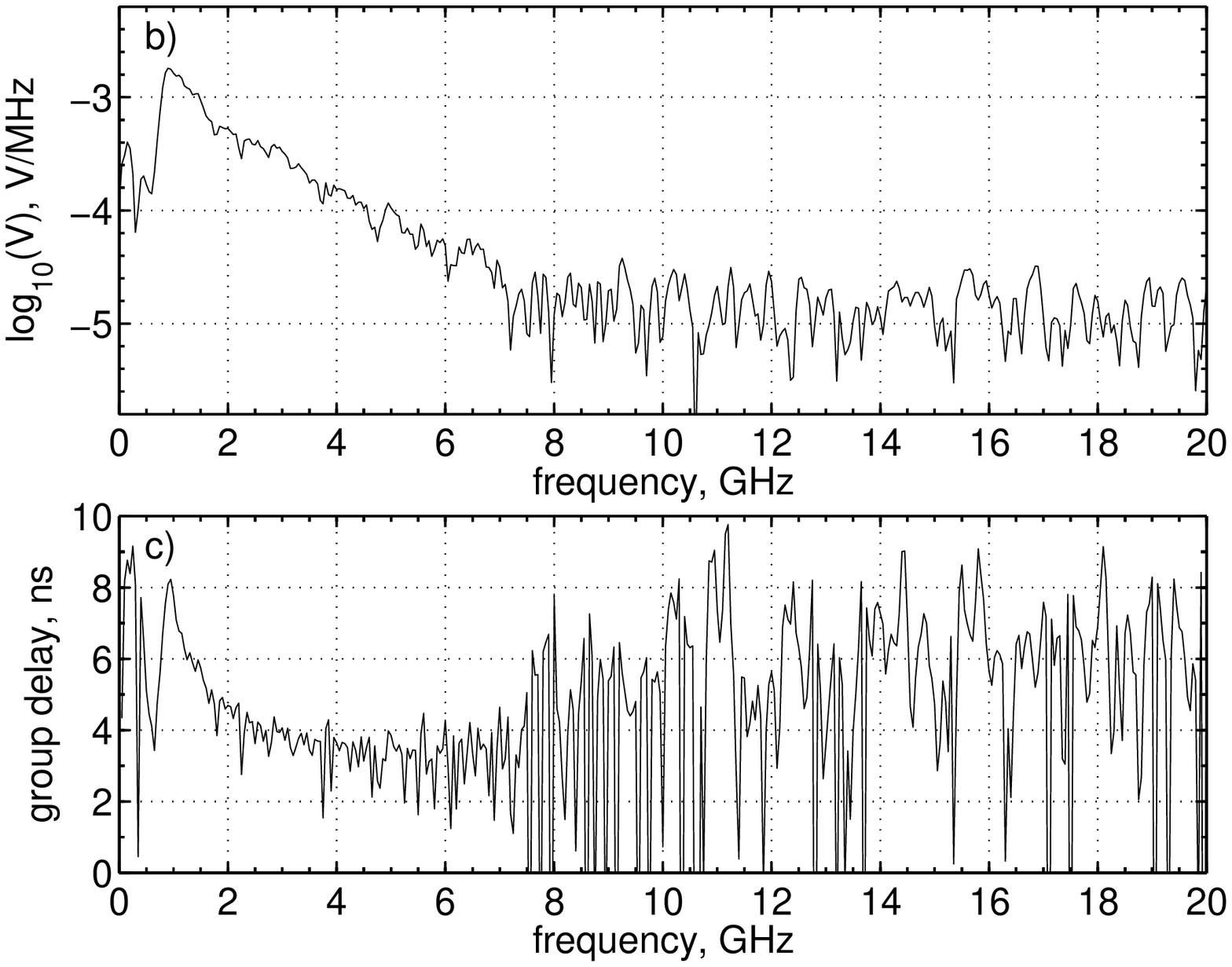}
\caption{(a) Raw voltage recorded with the LPDA in run 109, and its (b) magnitude and (c) group delay as 
functions of frequency.
\label{fig:raw}}
\end{figure}
\begin{figure}[t]
\includegraphics[scale=0.45]{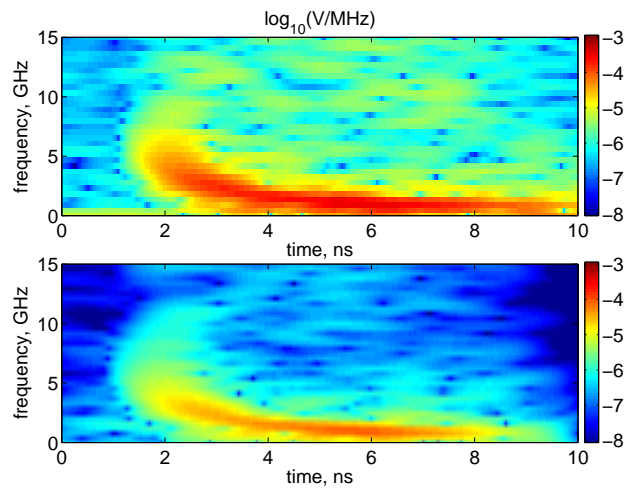}
\caption{(Color online) Spectrograms of signals recorded in runs 109 and 35 (top and bottom). The color indicates 
the logarithm of the square root of the signal power per unit frequency.
\label{fig:voltspec}}
\end{figure}

The raw signal recorded in run 109 is shown in Fig.~\ref{fig:raw}, along with its amplitude and group delay, defined as 
$\tau_g(\nu)=(2\pi)^{-1} \frac{d\phi(\nu)}{d\nu}$, where $\phi(\nu)$ is the relative phase of the voltage 
frequency component, $\phi=\omega t$.
It can be seen that the signal is present from $\sim$0.75~GHz (where the LPDA loses sensitivity) to $\sim$7.5~GHz, 
where the signal strength drops down to the intrinsic oscilloscope noise level. Run 35 signal looks identical except for 
the increased bandwidth (up to $\sim$11.5~GHz) since no attenuator was used. Fig.~\ref{fig:voltspec}
shows spectrograms of the recorded signals, with the curvature indicating variations in the delay of 
voltage frequency components due to the LPDA and the transmission line. 

\section{System Characterization}

In order to reconstruct the electromagnetic radiation pulse incident onto the LPDA (or any other antenna), the 
signal distorting effects of the antenna and the transmission line have to be corrected. The procedure outlined here 
is general and applicable to reconstruction of any observation of impulsive radio signals. 
The antenna equations and the required RF parameters are described in the appendix~\ref{sec:antennas}.

\subsection{Antenna system calibration}

The measurement of the LPDA effective height was performed in an anechoic chamber at University of 
Hawaii by a reciprocal S21 method. Two identical antennas were mounted about 60-in apart facing 
each other, which 
ensured they were in each other's far-field region. For an LPDA, the far-field requirement reduces 
to $d\ge\lambda/4$, where $d$ is the distance from the point at the LPDA where radiation of wavelength
$\lambda$ preferentially couples, {\it i.e.}, the phase center. The transmitting antenna was stimulated by a 
200--mV step impulse 
generated by an HP~54121A logging head. The received signal was amplified by an Agilent 83017A broadband 
amplifier and recorded by an HP~54120B digitizing oscilloscope with a 20 GHz bandwidth at 100 GSa/s. The presence 
of an amplifier introduces a transfer function to the system, so that the expression
for the recorded signal $V_{rec}$, combining Eqs.~\ref{eq:txrx} and~\ref{eq:tf}, is
\begin{equation}
\label{eq:calib}
V_{rec}(t) = \frac{1}{2\pi r c} H_{amp}(t) \circ h_N(t) \circ h_N(t) \circ \frac{d V_{src}(t)}{dt},
\end{equation}
where $r$ is the antenna separation, $c$ is the speed of light, $V_{src}$ is the voltage stimulating the transmitting
antenna, $h_N$ is the 
normalized effective height vector of the antennas, and $H_{amp}$ is the transfer function correction accounting 
for cables and the amplifier. The scalar product  (see Eq.~\ref{eq:txrx}) is omitted since the effective height 
vectors were parallel. The quantity $H_{amp}$ was measured by the same setup, but excluding the antennas from the 
circuit. Similarly, the heliax 
cable used in the SLAC measurement was stimulated on one side by a 200--mV step and the resulting pulse was 
recorded at the other end with the digitizing oscilloscope. The semi-rigid Haverhill cable was unavailable
for time-domain transfer function calibration, so only the attenuation was measured as a function of 
frequency with a network analyzer. Its phase response was ignored, but due to the relatively short length 
of that cable this omission will have a very small effect on the final result.  

\subsection{\label{sec:math}Mathematical operations}

In order to calculate the LPDA effective height from calibration measurements, it is conceptually
easiest to take a Fourier transform of Eq.~\ref{eq:calib} and re-arrange it so that
\begin{equation}
\label{eq:hnorm}
\tilde{h}_N(\nu)=\sqrt{\frac{r c}{i\nu}\frac{\tilde{V}_{rec}(\nu)}{\tilde{H}_{amp}(\nu) \tilde{V}_{src}(\nu)}},
\end{equation}
where $\tilde{f}(\nu)=\int f(t)e^{2\pi i \nu t}dt$.
Also, from Eq.~\ref{eq:tf}, $\tilde{H}_{amp}(\nu)=\tilde{V}_{out}(\nu)/\tilde{V}_{in}(\nu)$. 
The same expression is also applied to calculate the transfer function of heliax cable. All 
operations, including the square root, are performed using complex quantities. Special care should be taken when
taking a complex square root in order to properly account for phase wrapping. 

While straightforward, this frequency-domain based approach will give poor time-domain results if the frequency 
bandwidth of a measured signal is substantially narrower than the Nyquist limit of the sampling
oscilloscope. The time-domain result of a complex division operation in the frequency-domain will be dominated by frequency 
components where the signal is absent, {\it i.e.}, where a division of two small noise values can produce an arbitrarily 
large artificial result. Digital signal processing tools can be applied to filter out such artificial 
``out-of-band'' noise, but
they will in all cases produce an unacceptable level of signal distortion. In order to circumvent this
issue, a time-domain deconvolution algorithm with noise reduction can be applied in place of every
complex division operation. In this work, the Wiener algorithm was chosen~\cite{wiener}, and the level of 
the noise-to-signal power ratio required by the algorithm was chosen such to maximize out-of-band noise rejection 
while preserving the fidelity of the in-band signal (see Fig.~\ref{fig:ask_ant}b). 

\begin{figure}[t]
\includegraphics[scale=0.45]{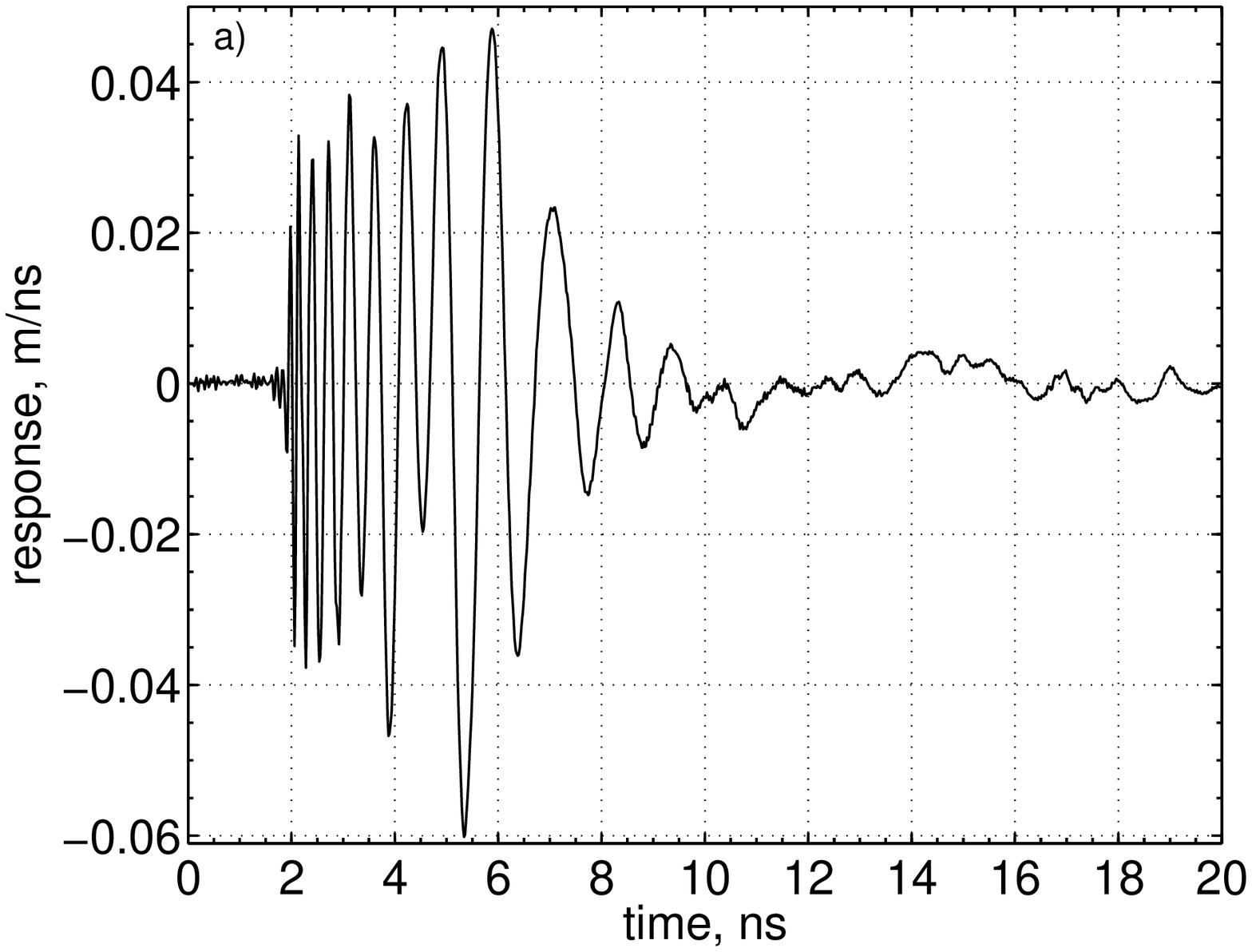}\\
\includegraphics[scale=0.45]{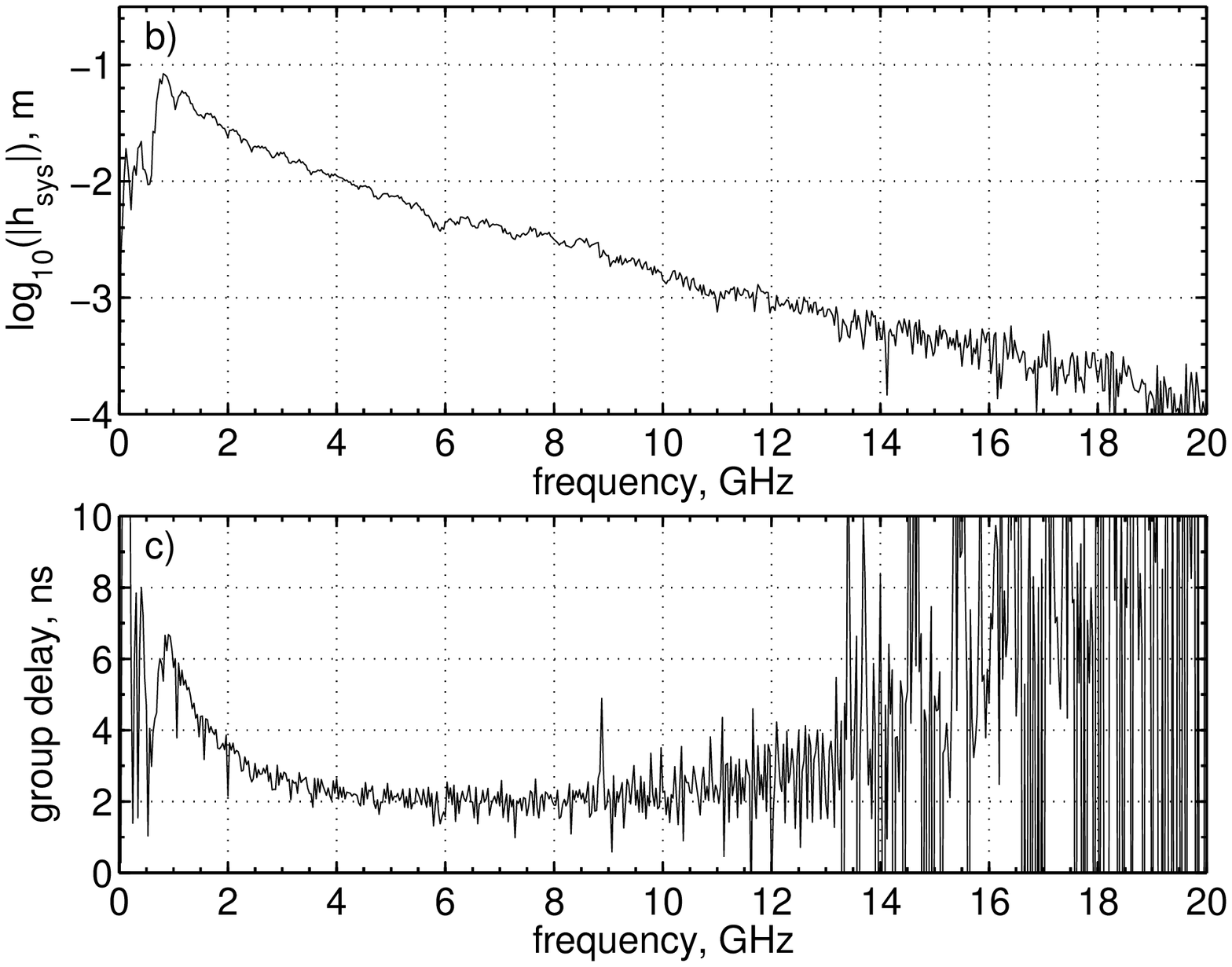}
\caption{(a) Time-domain response, and (b) attenuation and (c) group delay as functions of frequency for the LPDA-cable
system used in the measurements. The phase coherence of calibration pulses extends to 
$\sim$13~GHz, which exceeds the usable bandwidth of experimental data. The feature visible in group delay plot 
at $\sim$8.9 GHz is due to reflections at the amplifier input and it was not possible to calibrate it out. However, 
it does not create any appreciable effect in the final result.
\label{fig:calib}}
\end{figure}

The resulting time-domain response of the LPDA-cable system as used in the experiment can be 
expressed as $\bm{h}_{sys}(t)=H_{cable}(t) \circ \bm{h}_{N,LPDA}(t)$ and is shown in Fig.~\ref{fig:calib}.

\section{Analysis}

\begin{figure}[t]
\includegraphics[scale=0.45]{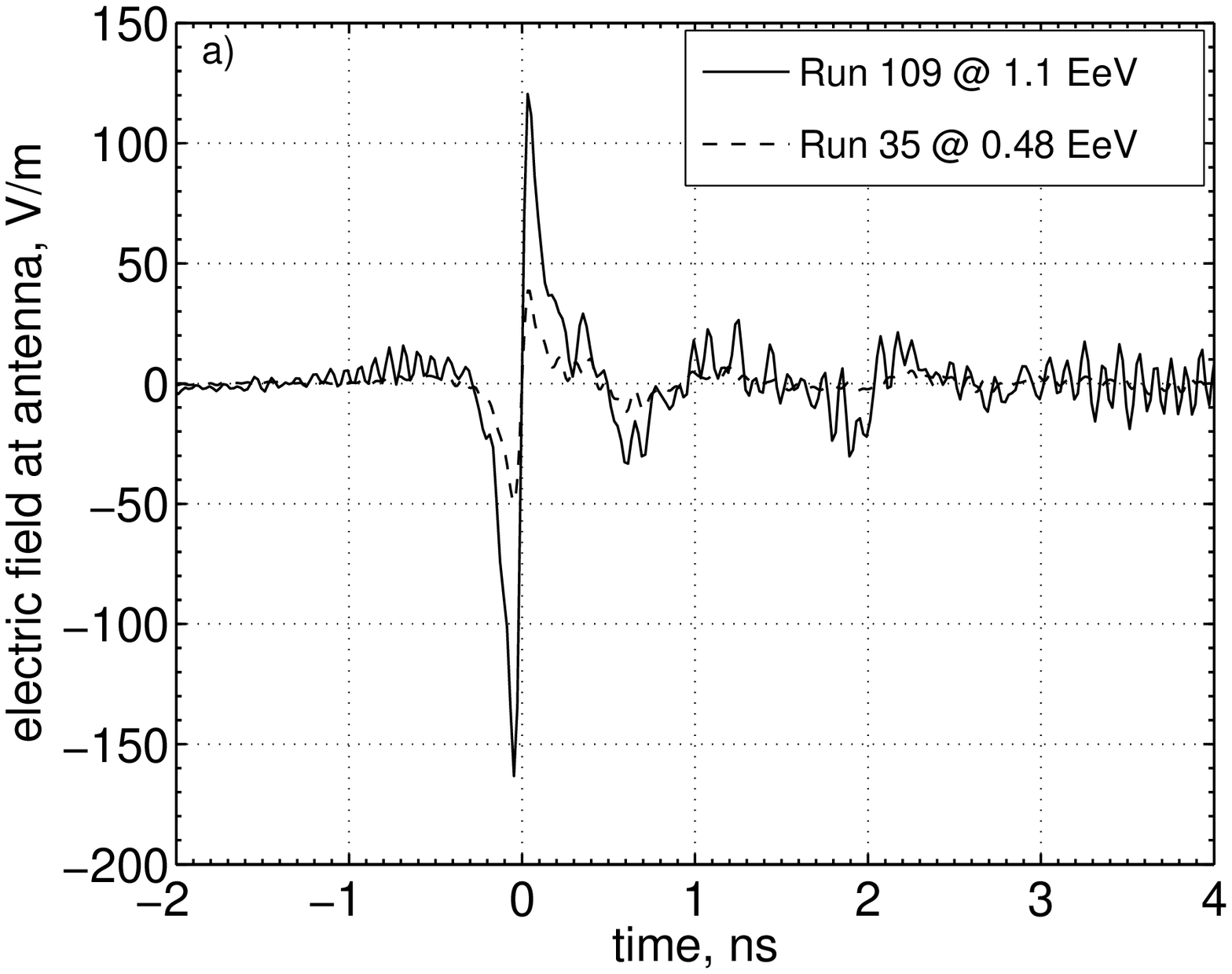}\\
\includegraphics[scale=0.45]{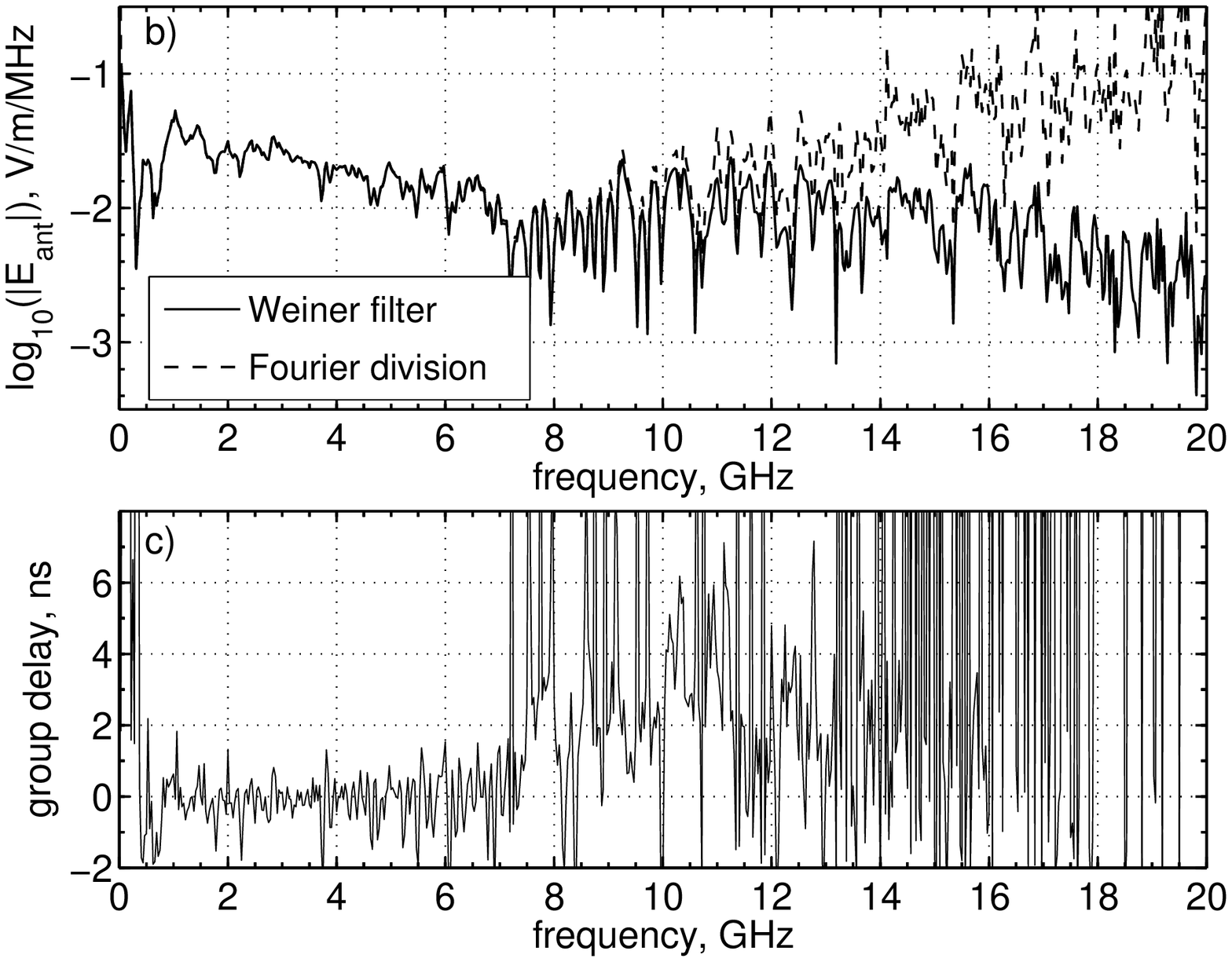}
\caption{The electric fields at the antenna in two runs (a), and the magnitude (b) and the group delay (c) as 
functions of frequency  (for run 109 only). In (b), the dashed line shows the magnitude obtained by straight 
frequency-domain division. The signal is dominated by artificial noise at frequencies outside of the signal 
bandwidth (0.75--7.5~GHz). It is clear 
that the use of the Wiener deconvolution algorithm is justified since it preserves the signal ``in-band'' while
suppressing the ``out-of-band'' noise. 
\label{fig:ask_ant}}
\end{figure}

The voltages recorded in the experiment can be expressed as (see Eqs.~\ref{eq:varsub2}--\ref{eq:rxnorm})
\begin{equation}
\label{eq:ask}
V_{rec}(t)=\sqrt{\frac{Z_L}{Z_0}} \bm{h}_{sys}(t) \star \bm{E}_{ant}(t),
\end{equation}
where $\bm{E}_{ant}$ is the electric field at the antenna due to the particle shower. During the 
measurements, the LPDA was aligned such that the polarization of electric field was parallel to the 
effective height vector, so that the spatial scalar product drops out of the equation. Thus, in 
the frequency domain, the electric field at the antenna is given by
\begin{equation} 
\label{eq:ant}
\tilde{E}_{ant}(\nu)=\sqrt{\frac{Z_0}{Z_L}}\frac{\tilde{V}_{rec}(\nu)}{\tilde{h}_{sys}(\nu)},
\end{equation}
with $Z_L=50\Omega$. The resulting electric field is shown in Fig.~\ref{fig:ask_ant}.  

Before the measured electric fields can be compared with the theoretical expectation, several corrections 
have to be applied. The theory was derived for a pulse detected at a very large distance from 
a shower initiated by a single particle, {\it i.e.}, where the shower can be considered a point source emitter 
and the radiation emitted is coherent over the full length of the shower. Additionally, the detection is also 
assumed to be in the same medium as the emission. In the experimental setup this is not the case. 
Aside from accounting for the electric field divergence as it crosses from salt to air, the portion of the 
particle 
shower generating a coherent pulse at the antenna scales with the frequency of observation. Finally, the number
of particles at the shower maximum, $N_{max}$, due to the superposition of many low-GeV electromagnetic showers 
will be larger than $N_{max}$ due to a shower initiated by a single particle of the equivalent energy. 
Since the intensity of the radiation emitted is proportional to $N_{max}$, this difference has to be taken
into account.

\subsection{Coherence zone correction}

\begin{figure}[t]
\includegraphics[scale=0.45]{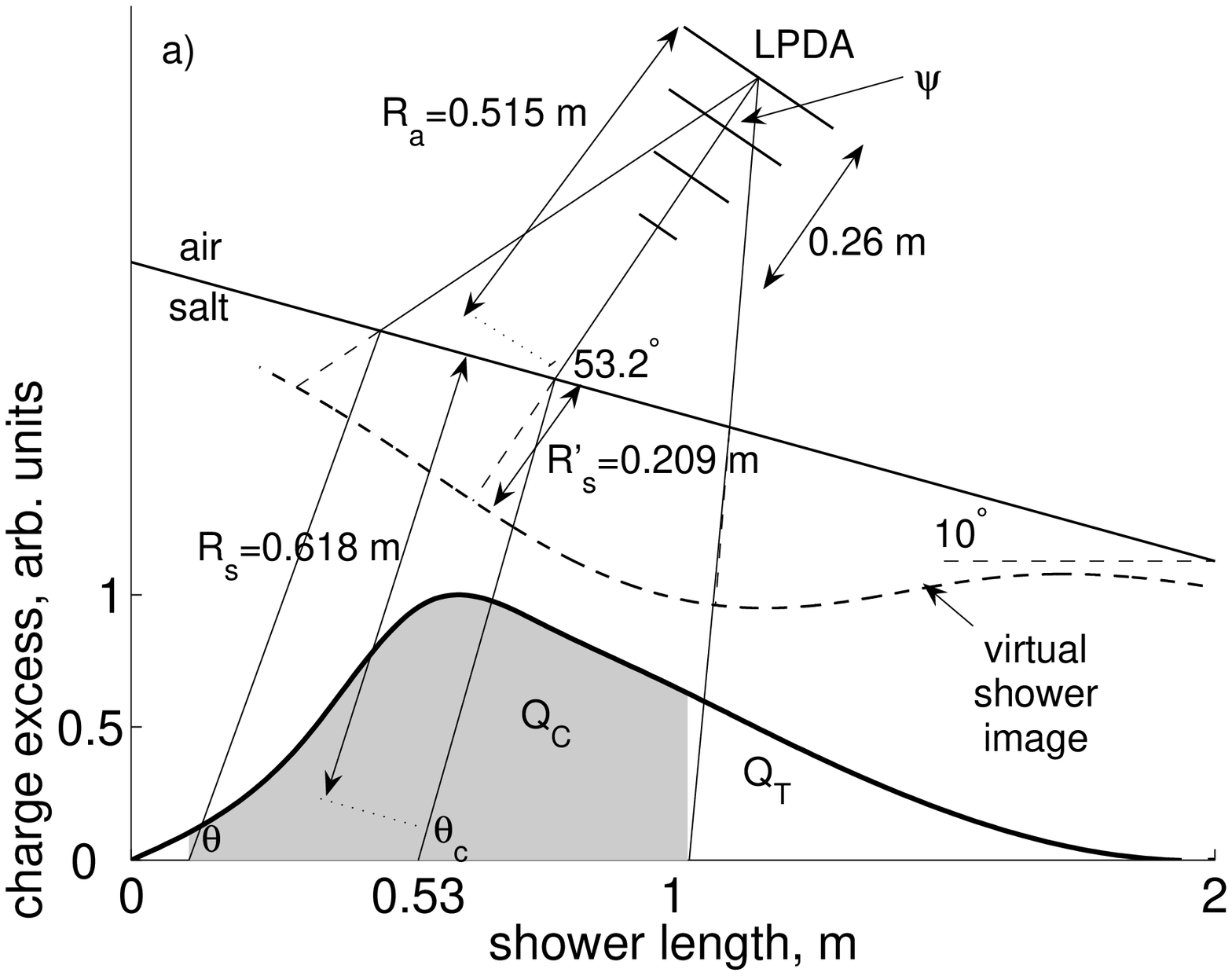}\\
\includegraphics[scale=0.45]{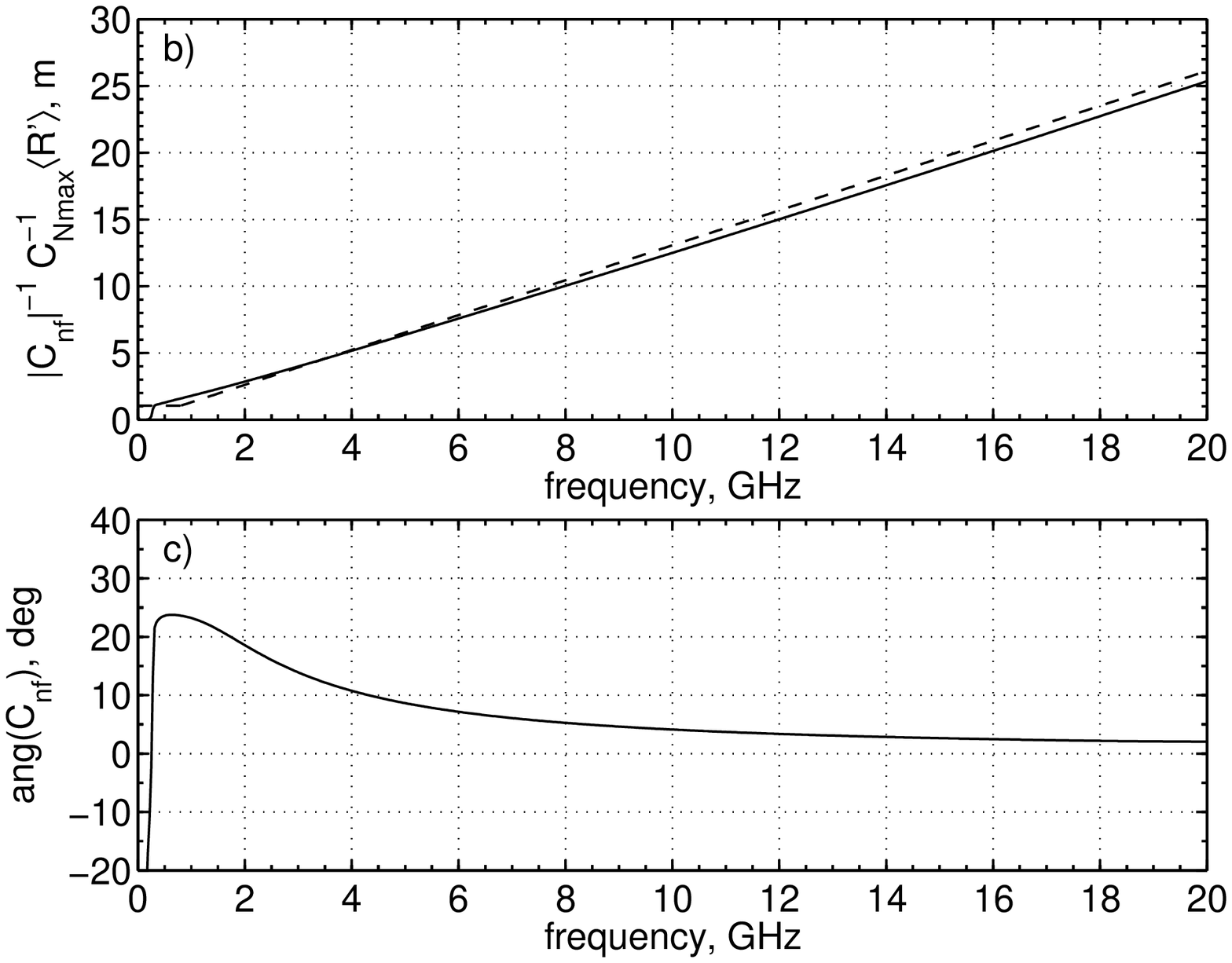}
\caption{a) A sketch of the experimental geometry, including the coherence zone (shaded) and the charge 
excess distribution as measured by the embedded bowtie antennas~\cite{slac05}. The dashed line, 
labeled virtual shower image, indicates the non-trivial path along which the shower appears to develop due
to the salt/air interface, as seen by an observer located at the LPDA feedpoint. 
b) The correction factor magnitude and c) phase as functions
of frequency (see text). The dashed line in b) is the {\it ad-hoc} correction factor used in Ref.~\cite{slac05}. 
\label{fig:raygraph}}
\end{figure}

The standard radiation coherence requirement is that phases of rays emerging from two ends of a coherence zone
do not differ by more than one cycle ($2\pi$) at any given frequency. The geometry of this requirement in
the present case is sketched in Fig.~\ref{fig:raygraph}a. 
Thus, the zone of coherence about a point on the
shower axis that the antenna is pointed at ($z_C$), can be defined for $z$'s satisfying the condition
\begin{equation}
\label{eq:dphi}
\Delta \phi(z;\nu) = \left|R''_s(z;\nu)-R''_s(z_C;\nu)+\frac{z-z_C}{n}\right| \frac{2\pi}{\lambda} \le \pi,
\end{equation}
where $z_C=0.53$~m, $n=2.44$ is the index of refraction of salt, radiation wavelength $\lambda=\frac{c}{n\nu}$,
and $R''_s(z;\nu)$ is the apparent distance from a point on the shower axis to 
the phase center of the antenna for the given frequency as if the entire path was in salt, {\it i.e.}, the number of actual 
phase cycles, $N$, is preserved and $R''_s=N\lambda$. Eq.~\ref{eq:dphi} is simply a restatement of 
the difference in phase factors of an electric field due to a moving charge without making a Fraunhofer 
zone approximation~\cite{jackson}. For the phase center of the
LPDA, a simple model is assumed where the phase center moves linearly along the antenna axis, such that the 
1--GHz phase center is at the feed-point and the 20--GHz phase center is at the tip of the antenna, about 26~cm from 
the feed-point. While not exact, this model is sufficiently accurate not to contribute strongly to 
systematic errors. As an example, the coherence zones at 1~GHz and 5~GHz were found to be $z=[0,1.05]$~m and 
$z=[0.45,0.77]$~m, respectively.

With the coherence zone defined, a correction to the electric field at the antenna with respect to the electric field at infinity 
can be made. From Refs.~\cite{zhs,am,forte}, the electric field due to coherent Cherenkov radiation at a given angle 
is proportional to the integral over the current distribution in the particle shower, $J(z)=c\;q(z)$, and to geometric factors 
relative to emission and observation points,
\begin{equation}
\label{eq:eprop}
E(\nu,\theta)\propto \int e^{i k(R\cos{\theta} + z/n)} q(z) dz,
\end{equation}
where $k=\frac{2\pi}{\lambda}$, $R$ is the distance from an emission point to the observation point, $\theta$ is 
the angle of emission, and the integral is taken over the full length of the shower. Taking into consideration 
the geometry of the experiment,\footnote{For ease of calculation and presentation, all geometry dependent 
quantities will be parametrized in terms of $z$, the distance along the shower axis, and $\nu$, 
the frequency.} a near-field correction factor relative to observation made at infinity along the Cherenkov 
angle can be defined as
\begin{equation}
\label{eq:cnf}
C_{nf}(\nu) = \frac{\int_{z_1(\nu)}^{z_2(\nu)} e^{i\Delta\phi} q(z)\;T\;G\;\frac{\sin{\theta}}{\sin{\theta_C}} e^{-\frac{1}{2}\left(\frac{\theta-\theta_C}{\Delta\theta(\nu)}\right)^2} dz}{\int_0^2 q(z) dz},
\end{equation}  
where $[z_1(\nu),z_2(\nu)]$ defines the coherence zone at a given frequency, $T\equiv T(z;\nu)$ is the 
transmission coefficient through the salt/polyethylene/air interface, $G\equiv G(z;\nu)$ is the relative
gain of the LPDA away from the antenna boresight, and $\theta\equiv\theta(z;\nu)$. 
The last two terms of the integrand in the numerator account for the drop in the electric field magnitude
away from Cherenkov angle, with $\Delta\theta=4.65^\circ\left[\frac{1\;\text{GHz}}{\nu}\right]$~\cite{am}. 

The transmission coefficient is given by the Fresnel equation for the E-field parallel to the plane of incidence~\cite{jackson},
\[
T=\frac{E'_{||}}{E_{||}}=\frac{2n\cos{i}}{n\cos{r}+n'\cos{i}},
\]
where $i$ and $r$ are incident and refracted angles related by Snell's law, and $n$ and $n'$ are indices of refraction in 
the two media. In calculating the transmission coefficient,
the presence of the 2.5~cm thick polyethylene sheet ($n_{poly}=2.25$) is ignored at frequencies below 
2.5~GHz (half-wavelength thickness) and its effect is gradually turned on to full strength above 10~GHz (two-wavelength 
thickness). The relative gain of the LPDA as a function of frequency was not measured, but was modeled based on 
the manufacturer's specification of $70^\circ$ average beam-width in the antenna E-plane\footnote{The plane parallel to $\bm{h}$.} 
and the standard LPDA expectation of first null at
twice the beam-width, to be $G(\psi)\approx\sqrt{\cos({\frac{90^\circ}{70^\circ}\psi})}$, with $\psi$ being the angle relative 
to the antenna boresight. 

When presenting the preliminary results of this analysis in Ref.~\cite{slac05}, it has been argued that the coherence zone
extends over the full shower at frequencies below 800~MHz, and that at higher frequencies the near-field correction
is proportional to $\nu$. In Fig.~\ref{fig:raygraph}b, it can be seen that this simplification was reasonable as it
produced an adequate near-field correction.

\subsection{Field divergence at the salt/air interface}

The $1/R$ attenuation in the electric field has to be modified for the field divergence at the salt/air interface. The 
field divergence factor for a spherical wave incident on a planar interface can be calculated by considering 
an area element $dA$ of the surface,
\begin{equation}
dA=\frac{R_s^2 d\Omega}{\cos{i}}=\frac{{R'_s}^2 d\Omega'}{\cos{r}},
\end{equation}
where $d\Omega$ is the solid angle element subtending $dA$ as seen from the emission point, $d\Omega'$ is the solid 
angle element subtending field lines exiting through $dA$, $R_s$ is the distance from the emission point to the salt/air 
interface, $R'_s$ is the distance from the virtual emission point as seen from the outside, and $i$ and $r$ are incident
and refracted angles as determined by Snell's law. The field divergence factor is given by the ratio of true and virtual 
distances~\cite{forte}, 
\begin{equation}
\label{eq:div}
\frac{R_s}{R'_s}=\frac{n\cos{i}}{\cos{r}}.
\end{equation}
The effective distance from an emission point to the LPDA is then given by,
\begin{equation}
\label{eq:Rp}
R'(z;\nu)=R_a(z;\nu)\frac{n\cos{i(z;\nu)}}{\cos{r(z;\nu)}}+R_s(z;\nu),
\end{equation}
where $R_a$ is the distance from the salt/air interface to the LPDA. The reader can easily convince oneself that the presence of 
the polyethylene sheet cancels out when considering the field divergence in the experimental setup. The mean LPDA to shower distance 
at any given frequency is then
\begin{equation}
\label{eq:Rpmean}
\langle R'(\nu)\rangle = \frac{\int_{z_1}^{z_2} R'(z;\nu) dz}{\int_{z_1}^{z_2} dz}.
\end{equation}

\subsection{$N_{max}$ correction}

The energy of photons initiating particle showers in the SLAC experiment was from $\sim$1--28500~MeV, with a $1/E$ 
bremsstrahlung distribution. The theoretical work on coherent Cherenkov radiation is based on simulations of 
electromagnetic showers initiated by primaries with energies of  
1~TeV and above~\cite{zhs,am}. While $N_{max}$, the number of particles at the shower maximum, scales nearly 
linearly with the energy of the primary at $E_p\agt10$~TeV, this is not the case for the lower energy primaries.
The EGS4 simulation shows that a bremsstrahlung photon bunch with the total energy of 20.7 TeV hitting the salt 
target will produce $N_{max}\approx55000$. However, the 20.7--TeV electron primary, interacting in the salt, 
gives rise to a shower whose peak is deeper and broader, and hence has only $\sim$19500 particles at the 
shower maximum~\cite{gaisser}.Thus, $C_{Nmax}=2.82$ can be defined as a correction factor to be used when 
comparing to theoretical results.

Putting all these corrections together, the measured electric field can be compared to the one which 
would have been expected from an equivalent shower observed in salt in the far-field region at Cherenkov angle, 
\begin{equation}
\label{eq:Efinal}
R|E(\nu,\theta_C)| = \frac{\langle R'(\nu)\rangle E_{ant}(\nu)}{C_{nf} C_{Nmax}}.
\end{equation}
The corrections will be applied only in the frequency range where the coherent signals are present in order not to
amplify the thermal noise background.

\section{Discussion}
     
\begin{figure}[bt]
\includegraphics[scale=0.45]{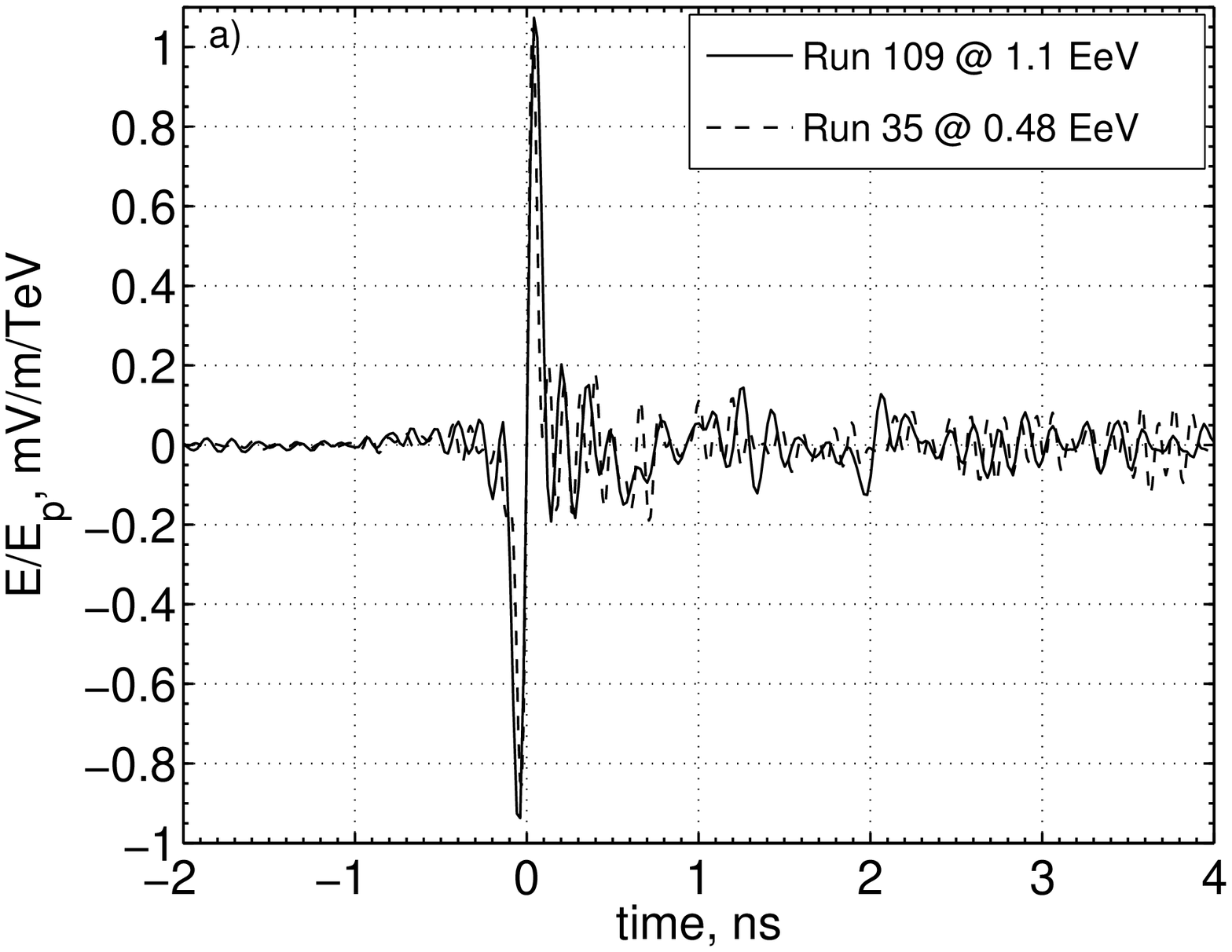}\\
\includegraphics[scale=0.45]{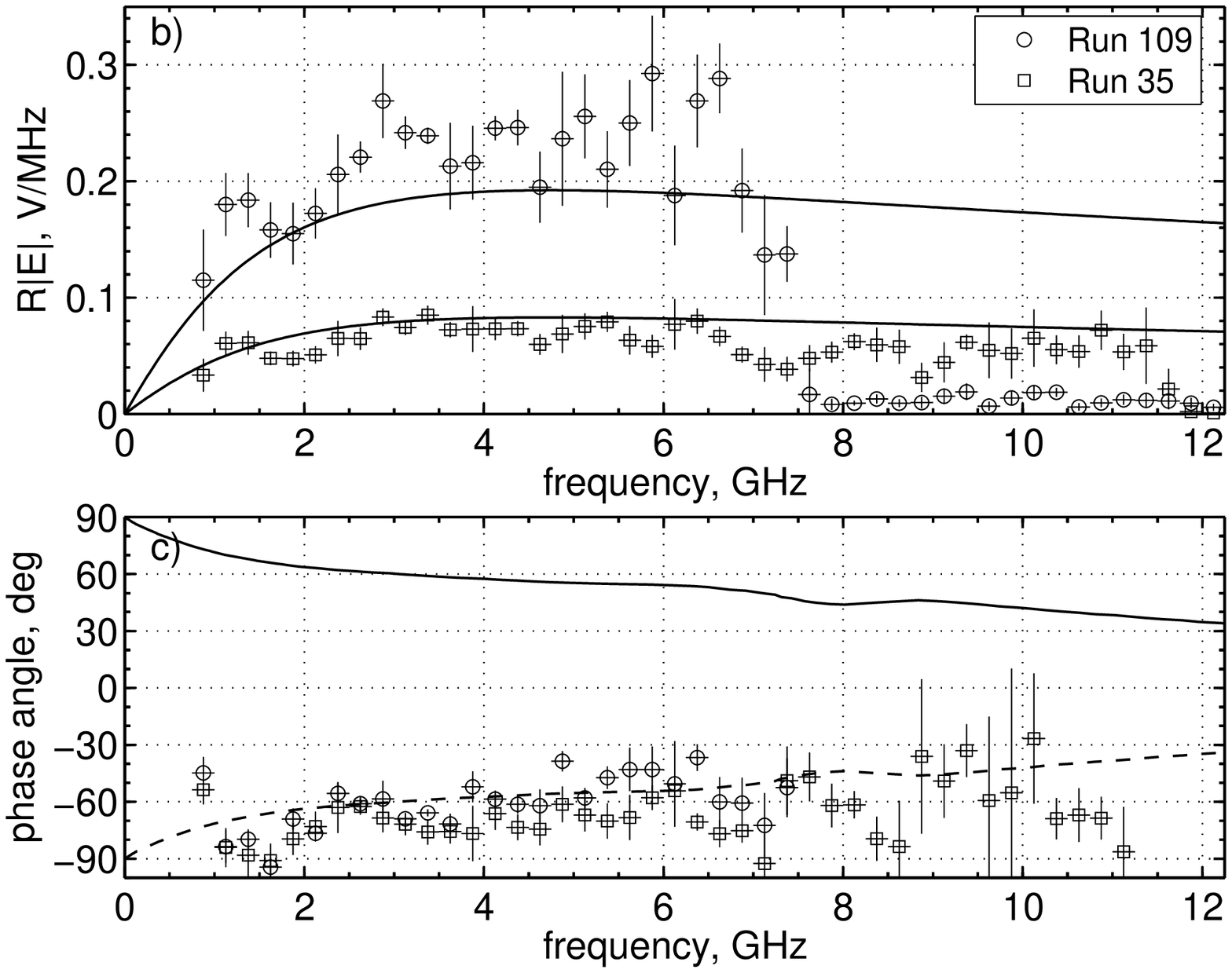}
\caption{(a) Electric field due to Askaryan pulses from two runs renormalized to 1~TeV primary energy; (b) electric field 
magnitudes as functions of frequency compared to the expectations; (c) electric field phases as functions of frequency, calculated 
relative to the zero-crossing point, compared to the expectation for an Askaryan pulse in salt (solid line)~\cite{zhs}. The dashed line 
places the theoretical phase expectation in the fourth quadrant, see text.
\label{fig:ask}}
\end{figure}

The far-field corrected pulses, and their magnitudes and phases, are shown in 
Fig.~\ref{fig:ask}. 
The magnitude and phase information have been binned in 250 MHz bins for clarity, and error bars indicate the RMS 
variation within each bin. The expectation for the magnitude of the electric field is given by~\cite{am,slac05} 
\begin{equation}
R|E|=A_0 f_d \left[\frac{E_p}{1 \text{TeV}}\right]\left[\frac{\nu}{\nu_0}\right]\left[\frac{1}{1+(\nu/\nu_1)^{1.44}}\right],
\end{equation}
where $A_0=2.53\times10^{-7}$~V/MHz, $f_d=0.52$, $\nu_0=1.15$~GHz, and $\nu_1=2.86$~GHz. The standard Fourier 
transform\footnote{$E(\omega)=\int_{-\infty}^{\infty}E(t)e^{i\omega t} dt$.} of the measured electric field was 
multiplied by a factor of 2 in order to agree with the definition of Eq.~8 in Ref.~\cite{zhs} adopted in 
Refs.~\cite{am,slac05}.

The only published electric field phase expectation has been calculated for an Askaryan pulse in ice~\cite{zhs}, but there is 
no reason to expect any significant deviation in the case of salt. However, there are two ambiguities that have to be 
resolved before a meaningful comparison of measured and expected phases can be made. The first is the actual quadrant of 
the phase. In the early theoretical calculations, the signs of the real and imaginary parts of the Fourier transform of the 
electric field were not kept~\cite{shahid}, and thus the complete phase information was unavailable. Newer calculations 
indicate that the phase is expected to lie in the fourth quadrant~\cite{shahid}, and the dashed line in Fig.~\ref{fig:ask}c 
indicates such a choice, which will be considered as correct in the remainder of this discussion. The second ambiguity is in the 
choice of the reference time about which to calculate the phase angle. No convention has been put forward, but one could 
expect that a natural choice would be the time of arrival of radiation from the shower maximum in the direction of the Cherenkov 
angle. Unfortunately, in the present experiment the relative oscilloscope trigger time with respect to the beam pulse was not 
recorded. The best guess would be to take the zero-crossing time in the middle of the pulse as was done in Fig.~\ref{fig:ask}c. 
A choice of a different reference time introduces an artificial phase slew, which can obscure the real physical features 
related to the phase, like the current distribution in the shower. 

\begin{figure}[t]
\includegraphics[scale=0.45]{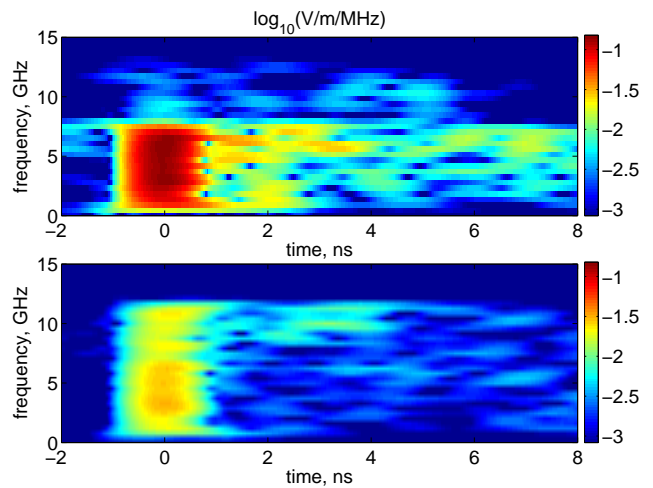}
\caption{(Color online) Spectrograms of electric fields due to Askaryan pulse recorded in runs 109 and 35 (top and bottom), where the color indicates the logarithm of the square root of electric field power per unit frequency. The reconstructed impulses are coherent and unresolved at all frequencies. 
\label{fig:spec}}
\end{figure}

A few features of the recorded pulses should be noted. The signals were recorded from 0.75--11.5 GHz in run 35 and 0.75--7.5 GHz
in run 109. This is indicated by the loss of both the phase coherence and the SNR above these frequencies due to attenuation 
in the antenna system, which increases with the frequency. As noted before, a 20~dB attenuator was present in run 109, 
causing the additional bandwidth loss. 
A clear way to present 
an impulsive, coherent, broadband signal expected from coherent Cherenkov radiation is with a spectrogram (Fig.~\ref{fig:spec}).
The pulse arrival is seen at zero time, with some low level reflections trickling in at the later times. 
The effect of deconvolving system response, Eq.~\ref{eq:ant}, and applying corrections, Eq.~\ref{eq:Efinal}, 
can be seen by comparing to Fig.~\ref{fig:voltspec}.

\begin{figure}[t]
\includegraphics[scale=0.42]{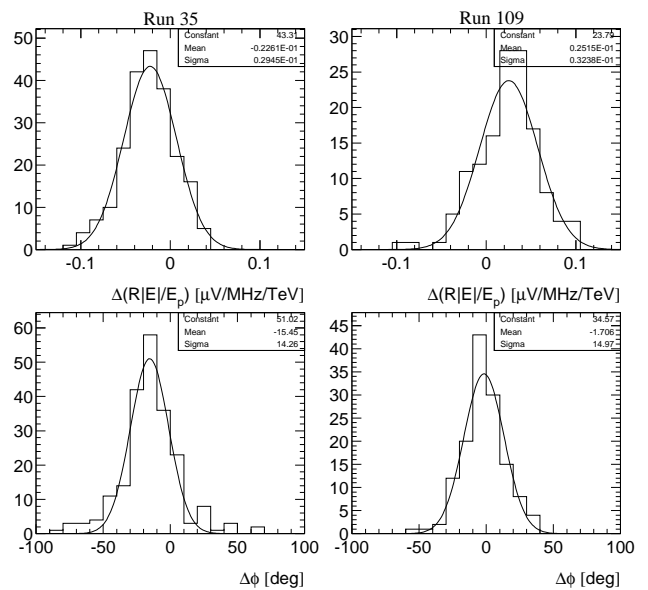}
\caption{Distributions of bin-by-bin differences in expected and measured magnitudes and phases of electric fields, over signal 
bandwidths noted in the text. The magnitudes have been renormalized to 1~TeV shower energy.  The fits are Gaussian.
\label{fig:diff}}
\end{figure}
Although deviations from theoretical expectations in Figs.~\ref{fig:ask}b and~\ref{fig:ask}c appear to be systematic since they are 
correlated between the runs, they are actually due to noise in the calibration data. This can be clearly seen in Fig.~\ref{fig:diff}
which histograms bin-by-bin differences between expectation and measurements over the signal bandwidth. The full frequency 
resolution of 50~MHz per bin was used. The deviations from expectation are Gaussian in nature, implying random noise. 
The offsets in the means of difference in magnitude can be attributed to systematic errors in the energy calibration of the runs, 
while the offset in the means of difference in phase, can be attributed to the choice of phase reference time. Changing the 
reference time by 8(1)~ps for run 35(109) shifts the mean to 0$^\circ$ while preserving the width of the distribution. 
Considering that the digitization time resolution is 10(20)~ps, this indicates a shift of less than one timing bin in both cases. 
Also, one could chose to use the predicted intensity of Cherenkov radiation in order to calibrate the total energy of showers
in two runs. Minimizing the mean of the difference in magnitude, the shower energies are found to be 0.38~EeV and 1.36~EeV, respectively. 
Combining the distributions from two runs, the fitted deviations are 
$\Delta(R|E|/E_p)=-0.005\pm0.039\text{(stat)}\pm0.045\text{(sys)}\;\mu\text{V/MHz/TeV}$ and
$\Delta\phi=-9^\circ\pm17^\circ\text{(stat)}\pm22^\circ\text{(sys)}$. Thus, it can be concluded that the theoretical expectations for 
magnitude and phase of the electric field~\cite{zhs,slac05} due to coherent Cherenkov radiation in salt in the frequency range from 
0.75--11.5~GHz have been confirmed within experimental uncertainty.

In this work, the procedure for the time-domain based analysis of antenna measurements and calibrations has 
been described and successfully applied to measurements of coherent Cherenkov radiation made at the FFTB 
facility at SLAC. The time-domain signal analysis preserves both magnitude and phase information of the 
original pulse, allowing for more accurate validation of theoretical models. The most precise validation  
of electric field intensity and the first validation of electric field phase of the theoretical model has been performed.

\appendix*
\section{Antenna equations}
\label{sec:antennas}

The voltage at the antenna port delivered to the transmission line of characteristic impedance $Z_L$ by the antenna with 
impedance $Z_{a,Rx}$ can be defined as~\cite{shliv,baum1},
\begin{equation}
\label{eq:rx}
V_L(t)  =  \tau_{Rx}\;\bm{h}_{e,Rx}(t) \star \bm{E}(t),
\end{equation}
with
\[
\tau_{Rx}  =  \frac{2 Z_L}{Z_{a,Rx}+Z_L},
\]
where $\star$ is the operator combining temporal convolution and spatial scalar product,\footnote{
$\bm{f}(t)\star\bm{g}(t)\equiv\int\bm{f}(t)\cdot\bm{g}(t-t')dt'$.} $\bm{E}$ is the incident electric field and 
$\bm{h}_{e,Rx}$ is the time-dependent effective height of the antenna for reception. The time-dependent effective 
height is the time-domain representation of the antenna complex impedance, and is equivalent to the antenna 
response to a delta-like impulse. 

At this point it is useful to also write down the equation for the electric field at some distance from
the antenna transmitting an impulsive signal driven through the transmission line of characteristic 
impedance $Z_L$ by a voltage source $V_{src}$, 
\begin{equation}
\label{eq:tx}
\bm{E}(t)  =  \frac{1}{4\pi r c} f_{Tx}\;\tau_{Tx}\;\bm{h}_{e,Tx}(t) \circ V_{src}(t), \\
\end{equation}
with
\begin{eqnarray*}
f_{Tx}  & =  & \frac{Z_0}{Z_{a,Tx}}, \\
\tau_{Tx} &  = & \frac{2 Z_{a,Tx}}{Z_L+Z_{a,Tx}}, 
\end{eqnarray*}
where $\circ$ is the convolution operator, $Z_{a,Tx}$ is the transmitting antenna impedance, $Z_0$ is the impedance of free space 
($377\Omega$), and $\bm{h}_{e,Tx}$ is the time-dependent effective height of the antenna for transmission. 
The voltage $V_{src}$ as used here is the voltage that would have been read by an oscilloscope if it 
were connected to the voltage source instead of an antenna and matched to the transmission line.\footnote{
Thus, $V_{src}$ is the voltage delivered to the antenna, not the open circuit voltage of the source.} 
Eq.~\ref{eq:tx} can be rewritten using the relation between transmission and reception 
effective heights of an antenna derived from self-reciprocity arguments~\cite{shliv},\footnote{The same conclusions based on self-reciprocity are derived by Baum~\cite{baum1}. Although it would appear that his result, ${\bm{h}}_{e,Tx}={\bm{h}}_{e,Rx}$ (using the notation adopted in this paper) is in disagreement with Ref.~\cite{shliv}, that is not the case since he uses slightly different definitions for various antenna system quantities.}
\begin{equation}
\label{eq:hrel}
\bm{h}_{e,Tx}(t)=2\partial_t\bm{h}_{e,Rx}(t).
\end{equation}
Substituting Eq.~\ref{eq:hrel} into Eq.~\ref{eq:tx} and noting that time derivation and convolution 
commute, the transmission equation can be written as, 
\begin{equation}
\bm{E}(t) = \frac{1}{2\pi r c} f_{Tx}\;\tau_{Tx}\;\bm{h}_{e,Rx}(t) \circ \frac{d V_{src}(t)}{dt}.
\end{equation}
Both antenna transmission and reception are now defined using the same effective height quantity. 

The form of the preceding equations assumes that both the antenna and the transmission line have purely
resistive impedance. The equations can be simplified and the complex antenna impedance can be 
reintroduced by absorbing the voltage transmission coefficient into the definition 
of effective height and renormalizing voltages and electric fields as suggested by Farr and 
Baum~\cite{farr}. Using the following variable substitutions,
\begin{eqnarray}
\label{eq:varsub1}
\bm{h}_N(t) & = & \frac{2 \sqrt{Z_0 Z_L}}{Z_a+Z_L}\bm{h}_{e,Rx}(t), \\
\label{eq:varsub2}
V_N(t) & = & \frac{V(t)}{\sqrt{Z_L}}, \\
\label{eq:varsub3}
\bm{E}_N(t) & = & \frac{\bm{E}(t)}{\sqrt{Z_0}},
\end{eqnarray}
where $Z_a$, $V$, and $\bm{E}$ stand for both transmission and reception cases, the antenna equations
become
\begin{eqnarray}
\label{eq:rxnorm}
V_{N,L}(t) & = & \bm{h}_N(t) \star \bm{E}_N(t), \\
\label{eq:txnorm}
\bm{E}_N(t) & = & \frac{1}{2\pi r c} \bm{h}_N(t) \circ \frac{d V_{N,src}(t)}{dt}.
\end{eqnarray}
Finally, if two identical antennas separated by distance $r$ are used to transmit and receive, and
transmission lines with the same impedance are used at both ends, the voltage observed on the port of 
the receiving antenna is given by
\begin{equation}
\label{eq:txrx}
V_L(t) = \frac{1}{2\pi r c} \bm{h}_N(t) \star \bm{h}_N(t) \circ \frac{d V_{src}(t)}{dt}.
\end{equation}

The transmission line complex impedance can be handled by introducing a transfer function correction, $H$,
which accounts for phase and amplitude distortions of a broadband signal traveling through the
transmission line. In the time-domain based description, this correction is expressed as
\begin{equation}
\label{eq:tf}
V_{out}(t) = H(t) \circ V_{in}(t).
\end{equation}
All of the antenna equations above need to include transmission line transfer function 
corrections at appropriate places, depending on the actual system setup. One should keep in mind
that convolution operators commute, so that contributions from several transmission line segments 
(even ones at the opposite transmission/reception ends of the system) can be combined into a single 
transfer function. 

\begin{acknowledgements}
This work was supported by the NASA ROSS - UH grant NAG5-5387 and DOE grant DE-FG02-04ER41291. The measurements 
were performed in part at the Stanford Linear Accelerator Center, under contract
with the US Department of Energy. 
\end{acknowledgements}

\end{document}